\documentclass[12pt]{article}
\usepackage{array}
\usepackage{graphicx}
\usepackage{amssymb}
\usepackage{amsmath}
\input{epsf}
\usepackage{cite}
\def\@fmsl@sh#1#2#3{\m@th\ooalign{$\hfil#1\mkern#2/\hfil$\crcr$#1#3$}}
 \def\eq#1\en{\begin{equation}#1\end{equation}}
\def\s[#1,#2]{[#1\stackrel{\star}{,}#2]}
\def\sx[#1,#2]{[#1\stackrel{\star_{x}}{,}#2]}

\newcommand{\nc}{\newcommand}
\nc{\beq}{\begin{equation}}
\nc{\eeq}{\end{equation}}
\nc{\beqa}{\begin{eqnarray}}
\nc{\eeqa}{\end{eqnarray}}

\def\bc{\begin{center}}
\def\ec{\end{center}}

\def\to{\rightarrow}

\def\gsim{\mathrel{\mathpalette\atversim>}}

\def\bc{\begin{center}}
\def\ec{\end{center}}

\def\gsim{\mathrel{\rlap{\lower4pt\hbox{\hskip1pt$\sim$}}

    \raise1pt\hbox{$>$}}}       

\def\gsim{\mathrel{\rlap{\lower4pt\hbox{\hskip1pt$\sim$}}
    \raise1pt\hbox{$>$}}}       

\begin{document}
\makeatletter
\def\fmslash{\@ifnextchar[{\fmsl@sh}{\fmsl@sh[0mu]}}
\def\fmsl@sh[#1]#2{%
  \mathchoice
    {\@fmsl@sh\displaystyle{#1}{#2}}%
    {\@fmsl@sh\textstyle{#1}{#2}}%
    {\@fmsl@sh\scriptstyle{#1}{#2}}%
    {\@fmsl@sh\scriptscriptstyle{#1}{#2}}}
\def\@fmsl@sh#1#2#3{\m@th\ooalign{$\hfil#1\mkern#2/\hfil$\crcr$#1#3$}}
\makeatother

\thispagestyle{empty}
\begin{titlepage}
\boldmath
\begin{center}
  \Large {\bf Asymptotically safe weak interactions}
  \end{center}
\unboldmath
\vspace{0.2cm}
\begin{center}
{
{\large Xavier Calmet}\footnote{x.calmet@sussex.ac.uk}
}
 \end{center}
\begin{center}
{\sl Physics and Astronomy, 
University of Sussex,  \\ Falmer, Brighton, BN1 9QH, UK 
}
\end{center}
\vspace{\fill}
\begin{abstract}
\noindent
We emphasize that the electroweak interactions without a Higgs boson are very similar to quantum general relativity. The Higgs field could just be a dressing field and might not exist as a propagating particle. In that interpretation, the electroweak interactions without a Higgs boson could be renormalizable at the non-perturbative level because of a non-trivial fixed point. Tree-level unitarity in electroweak bosons scattering is  restored by the running of the weak scale.
\end{abstract}  
\end{titlepage}



\newpage

The Anderson-Brout-Englert-Higgs-Guralnik-Hagen-Kibble mechanism, for short Higgs mechanism, represents an elegant way to generate masses for the electroweak gauge bosons while preserving the perturbative unitarity of the S-matrix and the renormalizability of the theory.
 It is however well understood that mass terms for the electroweak bosons can be written in a gauge invariant way using for example a non-linear sigma model representation \cite{Callan:1969sn} or a gauge invariant formulation of the electroweak bosons \cite{Calmet:2010cb}.  But,  if there is no propagating Higgs boson, quantum field theoretical amplitudes describing the scattering of the longitudinal modes of the electroweak bosons grow too fast with energy. In other words, perturbative unitarity is violated around a TeV \cite{LlewellynSmith:1973ey,Lee:1977yc,Lee:1977eg,Vayonakis:1976vz}. There are several ways unitarity could be restored \cite{Lee:1977yc}. Furthermore, the standard model without a Higgs boson is not renormalizable, at least at the perturbative level.
 
 One well studied possibility is that the weak interactions could become strongly coupled around a TeV. One then posits that the gauge theory unitarizes itself at the non-perturbative level. Another possibility for models without a Higgs boson consists in introducing weakly coupled new particles to delay the unitarity problem into the multi TeV regime where a UV completion of the standard model is expected to become relevant. Most recently it was proposed that, in analogy to black holes in gravitational scattering, classical objects could form in the scattering of longitudinal W-bosons leading to unitarized scattering amplitude \cite{Dvali:2010jz}. This mechanism is dubbed classification. 
 
 These different ideas are very interesting, they illustrate several facts about the electroweak standard model. First of all, the Higgs mechanism is not required to generate masses for the electroweak bosons, but its virtue consists in doing so in a way compatible with  perturbative unitary  and perturbative renormalizability. Secondly, the classification mechanism demonstrates that a theory can be unitary but not renormalizable. A consistent mechanism should thus fulfill three different and independent criteria: masses for the electroweak bosons need to be generated in a gauge invariant way, perturbative unitarity needs to be guaranteed and the theory needs to be renormalizable.

  In this short paper we propose an alternative approach to the unitarity and renormalization of the standard model without a Higgs boson which is based on a deep analogy between this model and general relativity. In both the non-linear sigma model version of the standard model and in its gauge invariant formulation, it is possible to define an action in terms of an expansion in the scale of the electroweak interactions $v$. The action of the model can be written schematically as
  \begin{eqnarray} \label{effaction}
S=S_{SM w/o Higgs}+\int d^4 x  \sum_i \frac{C_i}{v^n} O^{4+n}_i
\end{eqnarray}
 where $O^{4+n}_i$ are operators compatible with the symmetries of the model. The electroweak bosons are gauge invariant fields defined by
 \begin{eqnarray}
\underline W^i_\mu&=& \frac{i}{2g} \mbox{Tr} \ \Omega^\dagger \stackrel{\leftrightarrow}
{D_\mu} \Omega \tau^i
\end{eqnarray}
with $D_\mu=\partial_\mu - i  g  B_\mu(x)$ and 
  \begin{eqnarray}
\Omega=\frac{1}{\sqrt{\phi^\dagger
    \phi}}\left(\begin{array}{cc}  \phi_2^* & \phi_1 
    \\ -\phi_1^* & \phi_2
  \end{array}
\right )
\end{eqnarray}
 where 
 \begin{eqnarray}
\phi=\left(\begin{array}{c}  \phi_1 \\
   \phi_2
  \end{array}
\right ).
\end{eqnarray}
is a $SU(2)_L$ doublet scalar field which is considered to be a dressing field and does not need to propagate.  The very same construction can be applied to fermions \cite{Calmet:2010cb,tHooft:1998pk,'tHooft:1980xb,Mack:1977xu,Visnjic:1987pj}.

It is interesting to compare the effective action for the electroweak interactions (\ref{effaction})  to that of general relativity. General relativity is not renormalizable, at least at the perturbative level. The effective gravitational action at the quantum level is thus given by
\begin{eqnarray} \label{effaction2}
S[g]= -\int d^4x \sqrt{-\det(g)}\! \! \! \! \! && \left   (- \Lambda(\mu)^4+\frac{\bar M_P(\mu)^2}{32 \pi} R+  a(\mu) R_{\mu\nu} R^{\mu\nu} +b(\mu) R^2 
\right . \\ \nonumber && \left . +\frac{c(\mu)}{\bar M_P^2} R^3+\frac{d(\mu)}{\bar M_P^2} R R_{\mu\nu} R^{\mu\nu}+ .... \right ).
\end{eqnarray}
The analogy between the effective action for the electroweak interactions (\ref{effaction}) and that of quantum general relativity is quite striking. Both theories have a dimensionful energy scale which defines the strength of the interactions. The Planck mass sets this strength of the gravitational interactions while the weak scale determines the range and hence strength of the electroweak interactions. In the formulation proposed in \cite{Calmet:2010cb} where the Higgs field appears as a dressing field, the analogy between the weak interactions and the gravitational interactions can even be pushed further. In \cite{Calmet:2010cb}, the electroweak bosons are not gauge bosons. The local $SU(2)_L$ gauge symmetry is imposed at the level of the quantum fields. However there is a residual global $SU(2)$ symmetry, i.e. the custodial symmetry. In the case of general relativity, the tetrad is a necessary building block. It is a gauge field which transforms under local Lorentz
transformations $SO(3,1)$ and under general coordinate transformations, the metric $g_{\mu\nu}= e^a_\mu e^b_\nu \eta_{ab}$ which is the field that is being quantized, transforms under general coordinate transformations which is the equivalent of the global $SU(2)$ symmetry for the weak interactions. The close analogy between the tetrad field and the Higgs field is quite astonishing. It is very tempting to think of the Higgs field as the tetrad for the electroweak interactions while the electroweak bosons are similar to the metric.

While the gravitational action (\ref{effaction2}) is known to be non-renormalizable perturbatively, Weinberg proposed that the theory could be renormalizable at the non-perturbative level if there is a non-trivial fixed point \cite{fixedpoint}. This is the scenario of asymptotically safe gravity. This scenario implies that only a finite number of the Wilson coefficients in the effective action would need to be measured and the theory would thus be predictive.  It is rather tempting to propose that the weak interactions as defined by the effective action given in eq. (\ref{effaction}) have the same property. There are actually indications that a non-linear sigma model has a non-trivial fixed point \cite{Fabbrichesi:2010xy,Percacci:2009dt,Percacci:2009fh}. While for both the gravitational and electroweak interactions it is tremendously difficult to establish by calculations the existence of a fixed point, in the case of the electroweak interactions, this phenomenon will be probed in the coming years at the Large Hadron Collider. 

Measuring the strength of the electroweak interactions in the electroweak W-boson scattering could easily reveal a non-trivial running of the electroweak scale $v$. If an electroweak fixed point exists, experimentalists would first find an increase in the strength of the electroweak interactions, as in the strongly interacting W-bosons scenario, before the electroweak interactions become very weak and eventually irrelevant in the fixed point regime.  In analogy to the non-perturbative running of the non-perturbative Planck mass,  we introduce an effective weak scale
 \begin{eqnarray}
v_{eff}^2=v^2\left(1+\frac{\omega}{8\pi} \frac{\mu^2}{v^2} \right)
\end{eqnarray}
where $\mu$ is some arbitrary mass scale, $\omega$ a non-perturbative parameter which determines the running of the effective weak scale  and $v$ is the weak scale measured at low energies. If $\omega$ is positive, the electroweak interactions would become weaker with increasing center of mass energy. This is the asymptotically safe scenario for the weak interactions which would thus be renormalizable at the non-perturbative level without having a propagating Higgs boson again in complete analogy to Weinberg's proposal for quantum general relativity.

The asymptotically safe weak interaction scenario could also solve the unitarity problem of the standard model without a Higgs boson. 
In the standard model without a Higgs boson, there are five amplitudes contributing  at tree-level to the scattering of two longitudinally polarized electroweak W-bosons. Summing these five amplitudes, one finds at order $s/M_W^2$
\begin{eqnarray}
{\cal A}(W^+_L+W_L^-  \to W^+_L+W_L^-)= \frac{s}{v_{eff}^2} \left ( \frac{1}{2} + \frac{1}{2} \cos \theta \right )
\end{eqnarray}
where $s$ is the center of mass energy squared and $\theta$ is the scattering angle. Clearly if $v_{eff}$ grows fast enough with energy, the bad ultra-violet  of these amplitudes can be compensated and the summed amplitude can remain below the unitary bound. Note that a similar proposal has been made to solve problems with unitarity in models with large extra-dimensions \cite{Hewett:2007st}. Our scenario does not require any new physics beyond that already discovered. A careful monitoring of the strength of the electroweak interactions in the W-bosons scattering at the Large Hadron Collider could easily establish the existence of a fixed-point in the weak interactions. Let us discuss our idea further using the one-loop renormalization group of the weak scale \cite{Arason:1991ic}
\begin{eqnarray}
v(\mu)=v_0 \left (\frac{\mu}{\mu_0} \right)^\frac{\gamma}{16 \pi^2}
\end{eqnarray}
with
\begin{eqnarray}
\gamma=\frac{9}{4}\left ( \frac{1}{5} g_1^2 + g_2^2\right) -Y_2(S)
\end{eqnarray}
with
\begin{eqnarray}
Y_2(S)= \mbox{Tr} (3 \mbox{Y}_u^\dagger  \mbox{Y}_u + 3 \mbox{Y}_d^\dagger  \mbox{Y}_d + \mbox{Y}_e^\dagger  \mbox{Y}_e)
\end{eqnarray}
where $\mbox{Y}_i$ are the respective Yukawa matrices. Clearly as long as the theory is in the perturbative regime e.g. at $m_W$, the Yukawa coupling of the top dominates since at this scale $g_1=0.31$ and $g_2=0.65$ and $\gamma$ is negative. In other words, the scale of the weak interactions become smaller. However, if the weak interactions become strongly coupled in the TeV region, $g_2$ becomes large and $\gamma$ is expected to become positive and we obtain the expected running, i.e. the weak scale becomes larger. While we cannot prove using perturbation theory that there is a fixed point in the TeV region of the electroweak interactions, this result is encouraging for the scenario envisaged in this note. We stress once again that there are indications of a non-trivial fixed point for the non-linear sigma model using exact renormalization group techniques \cite{Fabbrichesi:2010xy,Percacci:2009dt,Percacci:2009fh}. The situation is thus very similar to that of quantum general relativity. In the case of the electroweak interactions this can be checked within the next few years at the Large Hadron Collider.

We have described here how the unitarity problem of the weak interactions could be fixed by a non-trivial fixed point in the renormalization group of the weak scale. A similar mechanism could also fix the unitarity problem for fermions masses \cite{Appelquist:1987cf,Maltoni:2000iq,Maltoni:2001dc,Dicus:2004rg,Dicus:2005ku} appearing if their masses are not generated by the Higgs mechanism.

Let us conclude by emphasizing that there are several proposals to avoid introducing a fundamental scalar degree of freedom in the standard model. Among Higgsless models, see e.g. \cite{Farhi:1980xs,Csaki:2003dt,Calmet:2008xe}, the closest to our proposal is that of Moffat \cite{Moffat:2010jx,Moffat:2008fd} which is based on a finite, non-local theory. It is however possible to differentiate both models experimentally. The model envisaged in this short paper has a very specific dynamics due to the fixed point in the electroweak interactions.




\bigskip


\end{document}